# An Adaptive Secret Key-directed Cryptographic Scheme for Secure Transmission in Wireless Sensor Networks

K. Muhammad[1], Z. Jan[2], J. Ahmad[3], Z. Khan[4]

[1,2,3,4,5]*Computer Science Department, Islamia College Peshawar, K. P. K., Pakistan*
[1]Khan.muhammad.icp@gmail.com

***Abstract-***Wireless Sensor Networks (WSNs) are memory and bandwidth limited networks whose main goals are to maximize the network lifetime and minimize the energy consumption and transmission cost. To achieve these goals, different techniques of compression and clustering have been used. However, security is an open and major issue in WSNs for which different approaches are used, both in centralized and distributed WSNs' environments. This paper presents an adaptive cryptographic scheme for secure transmission of various sensitive parameters, sensed by wireless sensors to the fusion center for further processing in WSNs such as military networks. The proposed method encrypts the sensitive captured data of sensor nodes using various encryption procedures (bitxor operation, bits shuffling, and secret key based encryption) and then sends it to the fusion center. At the fusion center, the received encrypted data is decrypted for taking further necessary actions. The experimental results with complexity analysis, validate the effectiveness and feasibility of the proposed method in terms of security in WSNs.

***Keywords-***Cryptography, Wireless Sensor Networks, Sensor Nodes, Fusion Center

## I. Introduction

A WSN consists of wireless sensor nodes that are capable to sense, compute, and communicate the data via a specific infrastructure [i]. WSNs are capable to monitor and track different activities and phenomenon that are difficult to monitor through human beings. For example, chemical environment monitoring, nuclear accident, and environment monitoring [ii-iv].

WSNs can be used to monitor different types of parameters including pressure, humidity, speed, temperature, presence of objects, lighting conditions, mechanical stress, direction, size of objects, and soil makeup. Some major and well-known applications of WSNs include under-water sensing, smart farming, forest fire detection, traffic monitoringand enforcement, anti-terrorism, target tracking, medical diagnosis, smart parking, multi-scale tracking, vineyard monitoring, image change detection, battle space monitoring by military, flood detection, networked gamming, remote sensing, habitat monitoring, smart video and audio surveillance, non-disruptive and nonintrusive monitoring of sensitive wildlife, and habitats [v-x]. Some applications of WSN in the field of engineering include civil structures monitoring, industrial plant maintenance, and modern building regulations using humidity and temperature [iii, vi, xi-xiii].

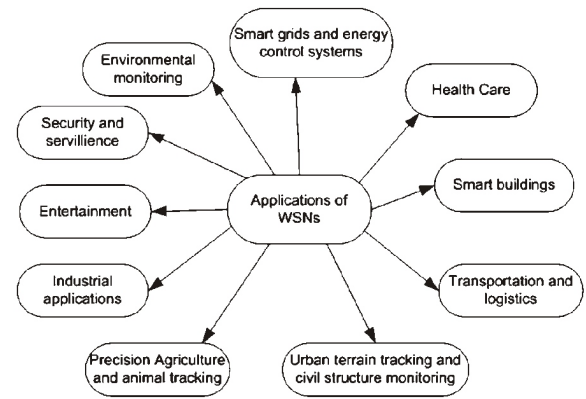

Fig. 1. Fields of Applications of WSN [xiv].

The most generic properties of WSN include limited bandwidth and energy, frequently changing topology, denser deployment of the network, multi-hop communication, autonomous management of the network, and limited transmission cost [xv]. The major issues and challenges that need to be addressed in WSN are network deployment (fixed/dynamic) depending upon the user requirements, network heterogeneity (less/more energy constraint), network scalability (Effect on the network due to node addition/deletion), uniform consumption of energy, communication model (single-hop/multi-hop), addressing based on attributes, and cluster dynamics (number of clusters, fixed/dynamic clusters, centralized/decentralized cluster head selection)[v-vi], [xvi-xviii].

### A. Emerging Areas of WSN

The recent emerging areas and their challenging issues that necessitate urgent solutions and invoke the researchers in diverse sophisticated multimedia applications of WSN are: [ii, xix]:

Image processing for network security in WSN

Localization of sensors based on images

Coverage of object view-angle using visual sensor networks

Object tracking using sensor image processing

Aggregation of images in sensor nodes

Image processing for minimizing the computations, bandwidth, and energy limitations

Pre-processing inside WSN like image compression

Efficient and effective capturing of video and images [xx]

*B.  Structure of a Typical Wireless Sensor Node*

A typical wireless sensor node consists of four core components and some optional components. The optional components depend on the nature of application. These components are briefly described below.

a. Each sensor node comprises a central processing unit plus some specific amount of memory for storage of intermediate results and other data, and a micro-controller.

b. Each sensor node occupies an explicit sensing unit that contains one or multiple sensors plus an A/D converter used for data attainment.

c. An RF unit used for communication of data using wireless media.

d. A special unit for providing power to sensor nodes.

e. A unit known as mobilizer for configuration and location changing. (optional component)

f. A system for position and location determination. (optional component)

This paper demonstrates a secure approach to handle the security issues in WSN by using cryptography and secret key. The proposed scheme increases the security of sensitive sensed data during transmission and avoids different fraudulent behaviors of adversaries. The main contributions of this paper are:

i. A new approach to handle the security issues in WSN using cryptography

ii. Encryption of sensitive sensed data using an adaptive cryptographic scheme where the level of encryption can be controlled by secret key, facilitating users to maintain a balance between security and available resources.

The remaining of the paper is structured as follows. Section 2 provides an overview of the classical and recent issues in WSN and their solutions whose major shortcomings let us toward current proposed work. The proposed cryptographic model is detailed in section 3. Section 4 is devoted to experimental results and discussion. Finally, section 5 concludes the paper.

## II. Related Work

Information security is a blooming research area.

Almost all the communicating bodies want confidentiality, secrecy, and integrity of their secret information [xxi]. Since the last decade, various classical and modern approaches have been proposed by researchers to handle these security issues. But still security is a major concern in this modern era of science and technology. WSNs, which are considered as open networks, are more vulnerable to different attacks and risks. Because of this reason, security issues in WSN have diverted the attention of researchers [xv, xvi].

According to [xxii] presented an energy efficient algorithm for image compression by making use of JPEG2000 compression scheme. This method increased the network life time and reduced the transmission cost. The major shortcomings of this approach are extra processing required by sensor nodes and its vulnerabilities to different diverse attacks and risks. The transmitted data in this scheme can be easily altered by hackers due to which the final decision taken by fusion center based on received data will be wrong, resulting in horrible destruction.

An area-based clustering detection (ABCD) technique is presented by reference [xxiii] to deal with the security issues of node replication problem in WSN environment. The said technique facilitates the users with high rate of correct detection and minimizes communication overhead as compared to line-selected multicast (LSM) approach. In contrast with centralized approach, the ABCD method minimizes the number of stored messages and extends the overall lifetime of the network.

In [xxiv], the authors present a brief discussion on the WSN security issues like integrity, confidentiality, authenticity, design and context related issues. The authors also nominated practical algorithm for data security and self-originating WSN for improving the performance and security properties in WSN.

Pathan, Lee, and Hong highlighted the foremost challenges and eminent attacks of WSN environment in [xxv]. According to them, the open challenges in WSN environment are accurate collection of data, secure data aggregation, trust management, and load balancing of resource constrained devices regarding their computation and communication. The authors critically discussed a number of attacks of WSN that invokes the WSN researchers for urgent solutions. Some of the possible attacks in WSN environment are denial of service, wormhole, hello flood, selecting forwarding, sinkhole, and Sybil attack. An advanced secure routing method was presented by reference [xxvi] for gray-hole attacks and false reports detection based on statistical en-route filtering for improving the security in WSN. Furthermore, energy consumption minimization and improved security of sensitive data is achieved using elliptical curve cryptography during its transmission.

The techniques discussed in previous paragraphs provide a single layer of security to the sensed data of wireless sensors during its transmission towards base station. To handle this issue, we propose a new approach with multiple levels of security, ensuring the secrecy of data during transmission.

## III. PROPOSED METHOD

In this section, the detailed description of the proposed method is presented. The proposed method handles the security issues in WSN environment, using secret key and cryptography, making the transmission of sensed sensitive data secure from different fraudulent behaviors. In most critical networks such as military systems, multi-scale tracking, and video surveillance systems, the tiny sensor nodes constantly sense the surrounding environment and transmit the sensed data using multi hop communication to the sink nodes. Thus, each node has to play two diverse roles: data gathering and performing as a rely point. It is then the responsibility of sink node to transmit the aggregated data to fusion center for further necessary processing and actions. This transmission of sensitive data is vulnerable to many risks and attacks.

The proposed approach handles these security flaws and issues during transmission of data from one node to another sensor node and lastly to fusion center in WSN environment with the help of multiple encryption algorithms including bitxor operation, bits shuffling and secret key based encryption. The diagrammatic representation of proposed cryptographic model is shown in Fig. 2.

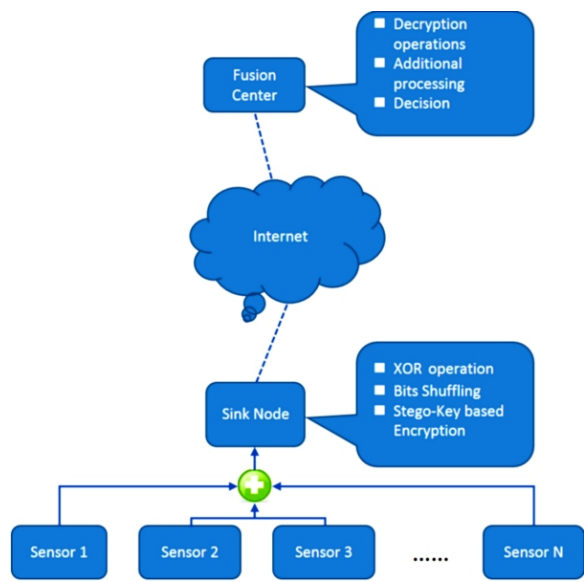

Fig. 2. The proposed cryptographic model for WSN security

The different types of sensors (audio sensors, video sensors, and scalar sensors) collect the most sensitive parameters from the surrounding specified serving area and send it to multimedia processing hub which aggregates the sensed parameters and send it to the sink node. At the sink, the aggregated sensitive secret information is encrypted via multiple encryption algorithms and is then transmitted to the base station. At the base station, the encrypted information is decrypted, some further processing is performed and an appropriate action is taken accordingly. Although, the proposed technique increases the processing of sensor nodes up to some extent but this is not a major issue for the most sensitive and critical sensor networks like atomic energy and intelligent agencies sensor networks because these networks cannot compromise on privacy and security. These reasons demonstrate that the proposed cryptographic model performs well in solving the security issues and problems in WSNs.

### A. Steps for Encryption

i. Convert the secret key into 1-D array of bits.

ii. Convert the sensitive sensed data of wireless nodes into 1-D array of bits.

iii. Shuffle all the key and message bits such that the bits with even and odd indices are interchanged.

iv. If SecretKeyBit (SKB)=1

          Then perform bitxor operation of secret message bit with logical 1.

    Else

          Do not perform bitxor operation.

    End if

v. Repeat step (iv) until all secret data bits are encrypted.

vi. Apply the bitxor operation on resultant bits with logical 1.

vii. Convert the resultant bits into its actual form.

### B. Steps for Decryption

i Take the appropriate secret key of particular sink node from the chunk of keys at the fusion center and convert it into 1-D array of bits.

ii Shuffle the resultant bits i.e. interchange the bits of odd and even positions with each other.

iii Convert the encrypted received information into bits form.

iv BitXOR the message bits with logical 1.

v If SecretKeyBit (SKB)=1

          Then perform bitxor operation of encrypted bit with logical 1.

    Else

          Do not perform bitxor operation.

    End if

vi Repeat step (v) until all bits are decrypted.

vii Shuffle the resultant bits by bits shuffling algorithm to get actual bits.

viii Convert the decrypted bits into actual data form.

## IV. EXPERIMENTAL SETUP AND DISCUSSION

The proposed cryptographic encryption and decryption algorithms are implemented using

MATLAB R2013a. For experiments some random numbers are taken as sensitive sensed data of sensor nodes and is encrypted and decrypted by the proposed technique. To make the idea easy to understand and avoid huge calculations, we have taken the secret data and secret key of limited length in the coming example. A complete case study of the proposed cryptographic model for security in WSN is shown in Table I and Table II.

Table I presents an example of encrypting the sensitive sensed data of sensor nodes. Two characters "A" and "B" are taken as sensed data and character "Z" is taken as a secret key. Col#5 (C5) shows the resultant bits after when the bits shuffling algorithm is applied on col#4 (C4) bits. In col#6 (C6), secret key based encryption algorithm is applied. If the key bit is 1, then the message bit is bitxored with logical 1, otherwise message bit remains unchanged. In col#7 (C7), the resultant bits are bitxored with logical 1 in order to further modify its shape and make the attack of malicious user awful. The final encrypted data is shown in col#8 (C8) which is transmitted by the sink node to fusion center for further processing and necessary actions.

TABLE I
EXAMPLE OF ENCRYPTION PROCEDURE; MB: MESSAGE BITS, KB: KEY BITS

| C1 | C2 | C3 | C4 | C5 | C6 | C7 | C8 |
|---|---|---|---|---|---|---|---|
| Secret sensitive data sensed by sensor nodes | Secret Key | ASCII Value | Binary Representation | Bits shuffling algorithm | Secret key based encryption algorithm | BitXOR operation with logical 1 | Encrypted data |
| - | Z | 90 | 01011010 | 10100101 | - | - | - |
| A | - | 65 | 01000001 | 10000010 | MB:10000010 KB:10100101 00100111 | 00100111 11111111 11011000 | Ø |
| B | - | 66 | 01000010 | 10000001 | MB:10000001 KB:10100101 00100100 | 00100100 11111111 11011011 | Û |

TABLE II
EXAMPLE OF DECRYPTION PROCEDURE: MB: MESSAGE BITS, KB: KEY BITS

| C1 | C2 | C3 | C4 | C5 | C6 | C7 | C8 |
|---|---|---|---|---|---|---|---|
| Received encrypted data | Secret Key | ASCII Value | Binary Representation | BitXOR operation with logical 1 | Secret key based decryption algorithm | Bits shuffling algorithm | Decrypted characters |
| - | Z | 90 | 01011010 | - | - | 10100101 | - |
| Ø | - | 216 | 11011000 | 11011000 11111111 00100111 | MB:00100111 KB:10100101 10000010 | 01000001 | A |
| Û | - | 219 | 11011011 | 11011011 11111111 00100100 | MB:00100100 KB:10100101 10000001 | 01000010 | B |

In Table II, the decryption procedure is briefly presented. The actual encrypted data received by base station is converted into bits form as shown in column 4 (C4) and is bitxored with logical 1 (col#5; C5). The secret key based decryption algorithm is applied on the resultant bits in col#6 (C6). At the end, the bits are shuffled with the bits shuffling algorithm (col#7; C7) and are converted to actual data form in col#8 (C8).

*A. Complexity analysis of the proposed method*

In this section, the complexity of the proposed method has been detailed. The complexity of the proposed technique depends on the length of secret key (key space) and the number of iterations during encryption. As mentioned in abstract that WSN are bounded in terms of network lifetime, bandwidth, and processing, therefore we have chosen a light-weight encryption algorithm to balance the security and processing overheads.

Currently, we have used a 64-bit key in actual simulation of the proposed method. The complexity of the proposed security system can be calculated as follows:

Key length=64 bits

Total number of possible keys= $2^{64}$

Suppose the attacker generates 1 million keys/second, then the total number of years required to break this algorithm using brute force attack will be as follows:

Number of keys generated/second=$10^6$

Total number of years required= $\dfrac{2^{64}}{10^6 \text{ x } 86400 \text{ x } 365}$

Average number of years required=292471

The proposed framework provides enough security to sensitive data during transmission. One can further increase the security by enlarging the key space but it can affect the performance in terms of processing time and memory consumption.

## V. Conclusions

In this paper, a new cryptographic model is proposed for coping with the security issues during transmission of sensitive data in WSNs. The different sensitive parameters sensed by tiny nodes are encrypted by an adaptive cryptographic scheme and are then transmitted to fusion center securely. Although, the anticipated technique requires a little bit more processing but in top-sensitive environments such as military and atomic energy sensor networks, this factor is acceptable as such departments cannot compromise on security. The proposed technique ensures the security of data during transmission and can be an excellent tool for adaptation of law enforcement agencies and military sensor networks for utilization in critical and security applications. Finally, it is concluded that the proposed scheme satisfies the favorable demands of current security systems with no extra transmission cost which confirms its superiority and effectiveness.

In future work, the authors will focus on the following key points to increase the WSN security up to a satisfied extent.

1. Encrypting the sensed data of sensor nodes with a more powerful encryption algorithm like RSA, Blowfish or DES.
2. Utilizing the concept of steganography to embed the sensed data inside a cover image for secure transmission and better security of WSN.
3. Designing an efficient and cost-effective algorithm to minimize transmission cost, power consumption, and processing.

## VI. Acknowledgment

The authors welcome the useful and constructive comments and prolific suggestions of the anonymous referees and area editors during the review process. Special thanks to Dr. Zahoor Jan, Mr. Jamil Ahmad, Mr. Haleem Farman, Mr. Zahid Khan, and anonymous referees whose valuable suggestions and encouraging comments improved the quality of this research work.